\RequirePackage{lineno}
\documentclass[aps,twocolumn,showpacs,byrevtex,prl,reprint]{revtex4-1}

\usepackage{graphicx}% Include figure files
\usepackage{dcolumn}% Align table columns on decimal point
\usepackage{bm}% bold math
\usepackage{rotating}
\usepackage{epstopdf}
\usepackage{color}
\usepackage{verbatim} %for comment
\usepackage{multirow}
\usepackage[abs]{overpic}
\usepackage{amsmath}
\usepackage{amssymb}
\usepackage{xspace}

\newcommand{\PreserveBackslash}[1]{\let\temp=\\#1\let\\=\temp}
\newcolumntype{C}[1]{>{\PreserveBackslash\centering}p{#1}}
\newcolumntype{R}[1]{>{\PreserveBackslash\raggedleft}p{#1}}
\newcolumntype{L}[1]{>{\PreserveBackslash\raggedright}p{#1}}

\newcommand{\EE}{e^+e^-}

\newcommand{\too}{\rightarrow}

\begin{document}
\graphicspath{{figure/}}
\DeclareGraphicsExtensions{.eps,.png,.ps}
\title{\boldmath Cross section measurement of $e^+e^- \rightarrow \eta'J/\psi$ from $\sqrt{s} = 4.178$ to $4.600$ GeV}
\author{
  \begin{small}
    \begin{center}
      M.~Ablikim$^{1}$, M.~N.~Achasov$^{10,d}$, P.~Adlarson$^{60}$, S. ~Ahmed$^{15}$, M.~Albrecht$^{4}$, M.~Alekseev$^{59A,59C}$, A.~Amoroso$^{59A,59C}$, F.~F.~An$^{1}$, Q.~An$^{56,44}$, Y.~Bai$^{43}$
% ...
\\
\vspace{0.2cm}
(BESIII Collaboration)\\
\vspace{0.2cm} {\it
$^{1}$ Institute of High Energy Physics, Beijing 100049, People's Republic of China\\
% ...
$^{4}$ Bochum Ruhr-University, D-44780 Bochum, Germany\\
% ...
$^{10}$ G.I. Budker Institute of Nuclear Physics SB RAS (BINP), Novosibirsk 630090, Russia\\
% ...
$^{15}$ Helmholtz Institute Mainz, Johann-Joachim-Becher-Weg 45, D-55099 Mainz, Germany\\
% ...
$^{43}$ Southeast University, Nanjing 211100, People's Republic of China\\
$^{44}$ State Key Laboratory of Particle Detection and Electronics, Beijing 100049, Hefei 230026, People's Republic of China\\
% ...
$^{56}$ University of Science and Technology of China, Hefei 230026, People's Republic of China\\
% ...
$^{59}$ (A)University of Turin, I-10125, Turin, Italy; (B)University of Eastern Piedmont, I-15121, Alessandria, Italy; (C)INFN, I-10125, Turin, Italy\\
$^{60}$ Uppsala University, Box 516, SE-75120 Uppsala, Sweden\\
% ...
$^{d}$ Also at the Novosibirsk State University, Novosibirsk, 630090, Russia\\
% ...
      }\end{center}
    \vspace{0.4cm}
\end{small}
}
\affiliation{}
%%% Local Variables:
%%% mode: latex
%%% TeX-master: "omega-chicj"
%%% End:

%\vspace{0.2cm}
%\date{\today}
%\linenumbers

\begin{abstract}
The cross section of the process $e^+e^- \rightarrow \eta'J/\psi$ is measured at center-of-mass energies from $\sqrt{s} =$ 4.178 to 4.600 GeV using data samples corresponding to a total integrated luminosity of 11 fb$^{-1}$ collected with the BESIII detector operating at the BEPCII storage ring. The dependence of the cross section on $\sqrt{s}$ shows an enhancement around $4.2$~GeV. While the shape of the cross section cannot be fully explained with a single $\psi(4160)$ or $\psi(4260)$ state,
a coherent sum of the two states does provide a reasonable description of the data.
\end{abstract}


\maketitle
%%%%%%%%%%%%%%%%%%%%%%%%%%%%%%%%%%%%%%%%%%%%%%%%%%%%%%%%%%%%%%%%%%%%%%%%%%%%%%
%%%%%%%%%%%%%%%%%%%%%%%%%%%%%%%%%%%%%%%%%%%%%%%%%%%%%%%%%%%%%%%%%%%%%%%%%%%%%%
\section{I. INTRODUCTION}
The Belle collaboration recently observed the transition $\Upsilon(4S)\too\eta'\Upsilon(1S)$~\cite{etapupsilon}.
It is therefore likely that a similar transition exists in the charmonium sector.
Moreover, CLEO-c, BESIII, and Belle measured the cross section as a function of $\sqrt{s}$ for the reaction $\EE \too \eta J/\psi$~\cite{cleo, bes, belle}, which apparently shows a significant contribution from $\psi(4160)$ decays.
In Ref.~\cite{theory}, the authors reproduce the measured $\EE \too \eta J/\psi$ line shape and predict the cross section of $\EE \too \eta' J/\psi$.
A measurement of the cross sections of $\EE \too \eta' J/\psi$ and $\eta J/\psi$ can thus help the development of related theories. The measured cross section of $\EE \too \eta'J/\psi$ can also be compared with that of $\EE \too \eta J/\psi$, which can provide more information to study charmonium(-like) states.
BESIII recently observed the process $\EE \too \eta'J/\psi$ using data collected at $\sqrt{s}$ = 4.226 and 4.258~GeV.
Due to limited statistics, no significant signal was observed at other energy values in the range from $4.189$ to $4.600$ GeV~\cite{etapjpsi}.
The line shape of the measured cross section could be reasonably described by a single $\psi(4160)$ state, supporting the hypothesis that the $\psi(4160)$ decays to $\eta'J/\psi$.
However, since the process $\EE \too \eta'J/\psi$ was only observed at two energy points, no conclusions could be drawn regarding possible additional states decaying to $\eta'J/\psi$.
Now that BESIII has collected more $\EE$ annihilation data samples around 4.2 GeV in 2016 and 2017, it is a good opportunity to search for the $\eta'$ transition $\psi(4160) \too \eta'J/\psi$ or $\psi(4260) \too \eta'J/\psi$, which will add another tile to our effort to understand the puzzle of the exotic states observed in the charmonium sector~\cite{omegachic0-bes, pipijpsi-bes, pipihc-bes, pipipsip-bes, piDDstar-bes}.

In this paper, we report a study of the reaction $\EE\too\eta'J/\psi$ based on the latest $\EE$ annihilation data collected with the BESIII detector~\cite{besiii} at fourteen energy points in the range 4.178 $\leqslant\sqrt{s}\leqslant$ 4.600 GeV, with a total integrated luminosity of about 11 fb$^{-1}$. The $\eta'$ state is reconstructed via $\eta' \too \gamma\pi^+\pi^-/\pi^+\pi^-\eta$~\cite{gammapipi}, $\eta\too\gamma\gamma$ decays, and the $J/\psi$ is reconstructed via $J/\psi \too \ell^{+}\ell^{-}$ ($\ell=e$ or $\mu$) decays.

\section{II. BESIII DETECTOR AND MONTE CARLO SIMULATION}
The BESIII detector is a magnetic spectrometer~\cite{besiii} located at the Beijing Electron
Positron Collider (BEPCII)~\cite{bepcii}. The
cylindrical core of the BESIII detector consists of a helium-based
 multilayer drift chamber (MDC), a plastic scintillator time-of-flight
system (TOF), and a CsI(Tl) electromagnetic calorimeter (EMC),
which are all enclosed in a superconducting solenoidal magnet
providing a 1.0~T magnetic field. The solenoid is supported by an
octagonal flux-return yoke with resistive plate counter muon
identifier modules interleaved with steel. The acceptance of
charged particles and photons is 93\% over the $4\pi$ solid angle. The
charged-particle momentum resolution at $1~{\rm GeV}/c$ is
$0.5\%$, and the $\textrm{d}E/\textrm{d}x$ resolution is $6\%$ for the electrons
from Bhabha scattering. The EMC measures photon energies with a
resolution of $2.5\%$ ($5\%$) at $1$~GeV in the barrel (end cap)
region. The time resolution of the TOF barrel part is 68~ps. The end cap TOF
system was upgraded in 2015 with multi-gap resistive plate chamber
technology, providing a time resolution of
60~ps~\cite{etof}.

Simulated data samples produced with the {\sc geant4}-based~\cite{geant4} Monte Carlo (MC) package, which
includes the geometric description of the BESIII detector and the
detector response, are used to determine the detection efficiency
and to estimate the background contributions. The simulation includes the beam
energy spread and initial-state radiation (ISR) in the $e^+e^-$
annihilations modeled with the generator {\sc kkmc}~\cite{KKMC}.
Signal MC samples for $\EE \too \eta'J/\psi$ are generated at each center-of-mass energy point assuming that the cross section follows a coherent sum of a $\psi(4160)$ Breit-Wigner (BW) function and a $\psi(4260)$ BW function, with masses and widths are fixed to their Particle Data Group (PDG) values~\cite{pdg}.
The inclusive MC samples consist of the production of open charm
processes, the ISR production of vector charmonium(-like) states,
and the continuum processes incorporated in {\sc kkmc}~\cite{KKMC}. The known decay modes are modeled with {\sc evtgen}~\cite{ref:evtgen} using branching fractions summarized and averaged by the
PDG~\cite{pdg}, and the remaining unknown decays
from the charmonium states are generated with {\sc lundcharm}~\cite{ref:lundcharm}. Final state radiation from charged final state particles is incorporated with the {\sc photos} package~\cite{photos}.

\section{III. EVENT SELECTION}
For each charged track, the distance of closest approach to the interaction point (IP) is required to be within $10$ cm in the beam direction and within 1 cm in the plane perpendicular to the beam direction. The polar angles ($\theta$) of the tracks must be within the fiducial volume of MDC $(|\cos\theta|<0.93)$. Photons are reconstructed from isolated showers in EMC, which are at least $20^\circ$ away from the nearest charged track. The photon energy is required to be at least 25 MeV in the barrel region $(|\cos\theta|<0.8)$ or 50 MeV in the end cap region $(0.86<|\cos\theta|<0.92)$. To suppress electronic noise and energy depositions unrelated to the event, the EMC cluster timing from the reconstructed event start time is further required to satisfy $0\leq t \leq 700$ ns.

Since the reaction $\EE \too \eta'J/\psi$ results in the final states $\gamma\gamma\pi^{+}\pi^{-}e^+e^-/\mu^+\mu^-$ and $\gamma\pi^{+}\pi^{-}e^+e^-/\mu^+\mu^-$, candidate events are required to have four tracks with zero net charge, at least two good photons for $\eta' \too \pi^+\pi^-\eta$, and at least one for $\eta' \too \gamma\pi^+\pi^-$.
Tracks with momenta larger than 1~GeV/$c$ are assigned as leptons from the decay of the $J/\psi$; otherwise, they are considered as pions from $\eta'$ decays.
Leptons from the $J/\psi$ decay with energy deposited in EMC larger than 1.0~GeV are identified as electrons, and those less than 0.4~GeV as muons.
To reduce the background contributions and to improve the mass resolution, a four-constraint (4C) kinematic fit is performed for the $\eta' \too \gamma\pi^+\pi^-$ decay mode, constraining the total four-momentum of the final state particles to the total initial four-momentum of the colliding beams.
A five-constraint~(5C) kinematic fit is
performed for the $\eta' \too \pi^+\pi^-\eta$ decay mode both to constrain
the total four-momentum of the final state particles to the total initial four-momentum of the colliding beams and to constrain the invariant mass of the two photons from the decay of the $\eta$ to its nominal mass~\cite{pdg}.
If there is more than one combination in an event, the one with the smallest $\chi^{2}_{\text{4C}}$ or $\chi^{2}_{\text{5C}}$ of the kinematic fit is selected.
The $\chi^{2}_{\text{4C}}$ or $\chi^{2}_{\text{5C}}$ of the candidate events is required to be less than 40 or 50, respectively.

Besides the requirements described above, further selection criteria  are applied. For the decay channel $\eta' \too \pi^+\pi^-\eta$, in order to eliminate background from $\EE\too\pi^+\pi^-\psi(2S)\too\pi^+\pi^-\eta J/\psi$, the $\eta J/\psi$ invariant mass $M(\eta J/\psi)$ is required to be outside the region $(3.67, 3.70)$~GeV/$c^2$.
For the decay channel $\eta' \too \gamma\pi^+\pi^-$, in order to remove background from $\EE\too\gamma_{ISR}\psi(2S)\too\gamma_{ISR}\pi^+\pi^-J/\psi$, the invariant mass $M(\pi^+\pi^-J/\psi)$ is required to be outside the region $(3.66, 3.71)$~GeV/$c^2$, and in order to remove background from photon conversions, the cosine of the angle between the $\pi^+$ and $\pi^-$, cos$\theta_{\pi^+\pi^-}$, is required to be less than 0.95.

\section{IV. BORN CROSS SECTION MEASUREMENT}
\begin{figure}[htbp]
\begin{center}
\begin{overpic}[width=0.23\textwidth]{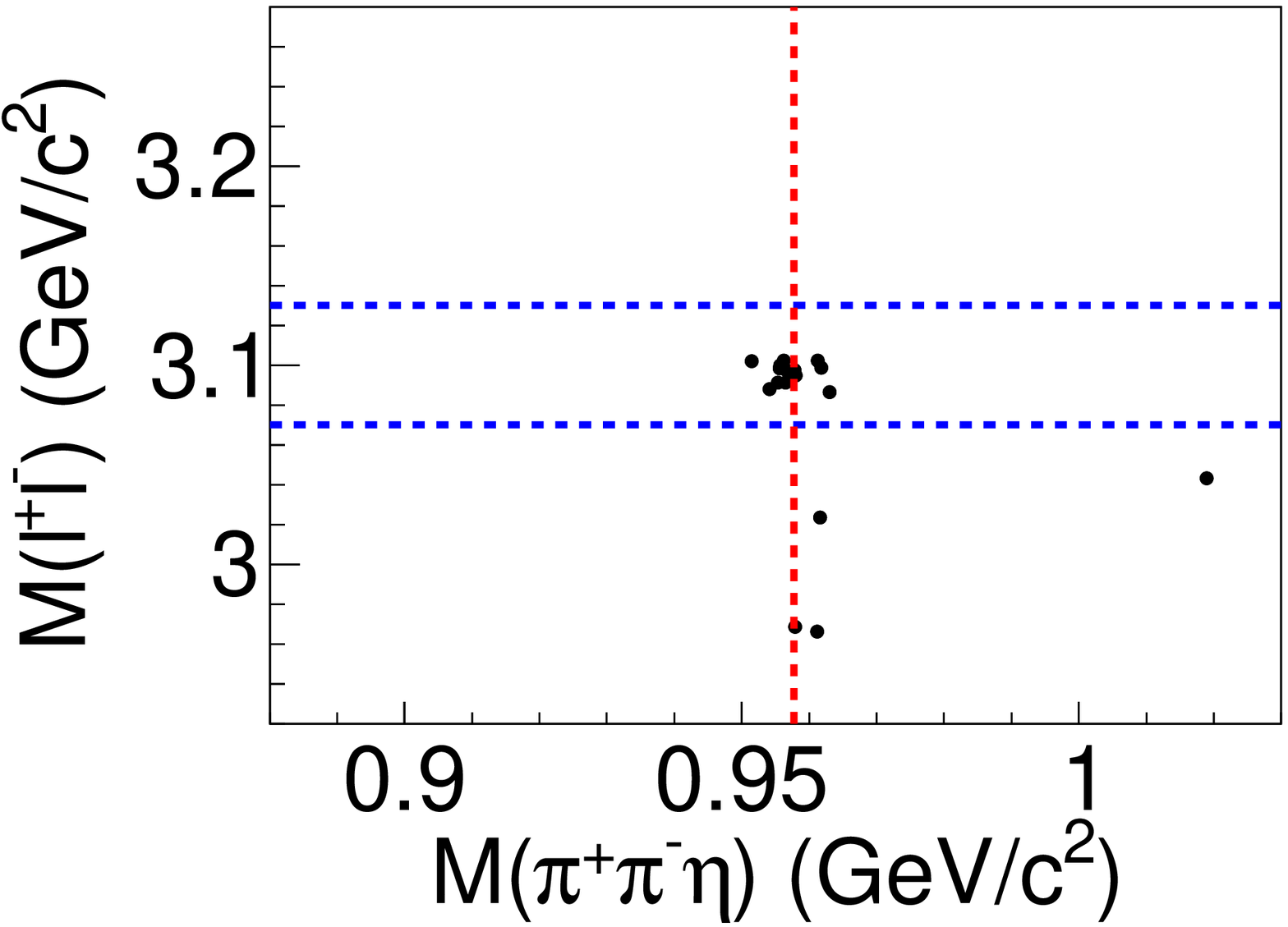}
\put(97,69){(a)}
\end{overpic}
\begin{overpic}[width=0.23\textwidth]{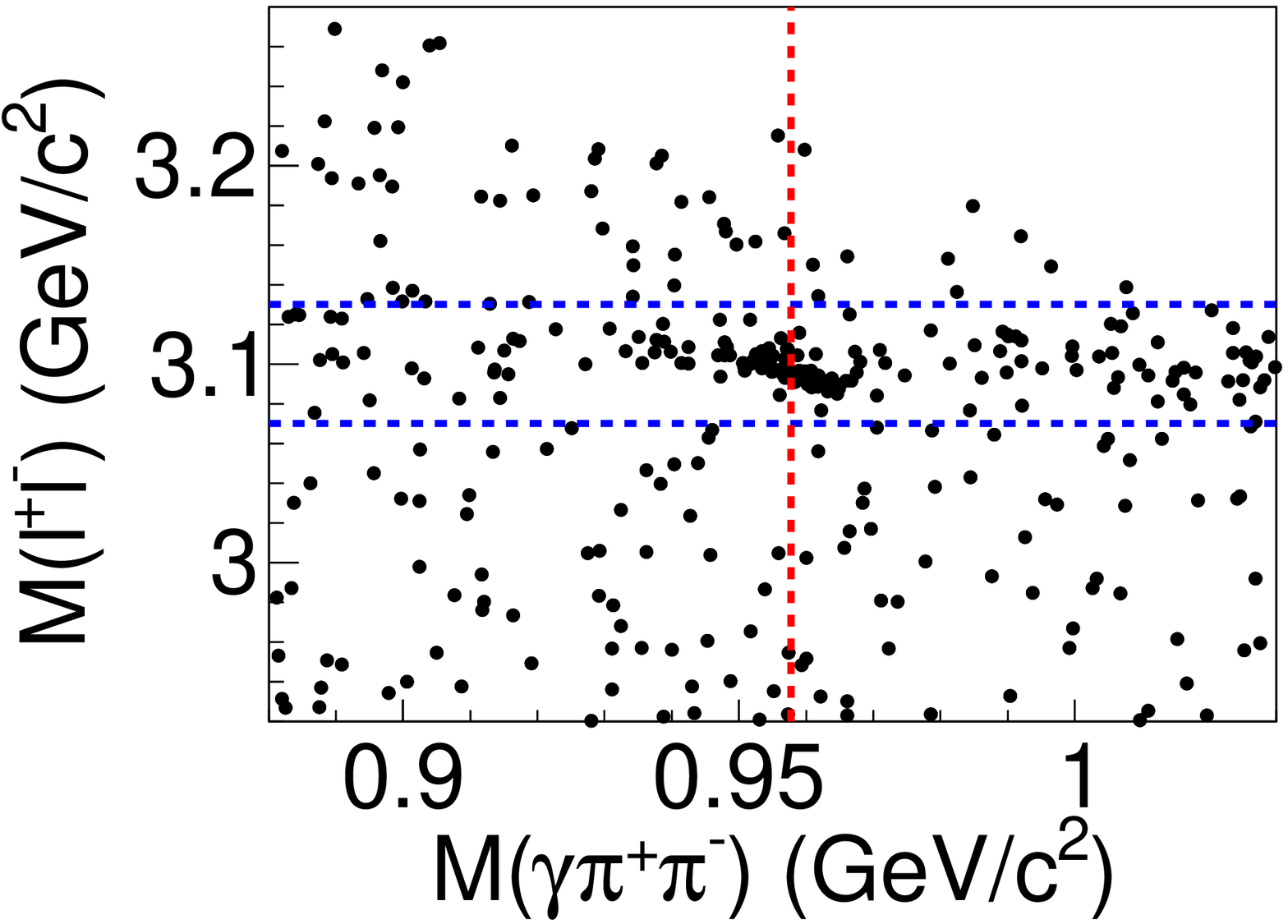}
\put(97,69){(b)}
\end{overpic}
\begin{overpic}[width=0.23\textwidth]{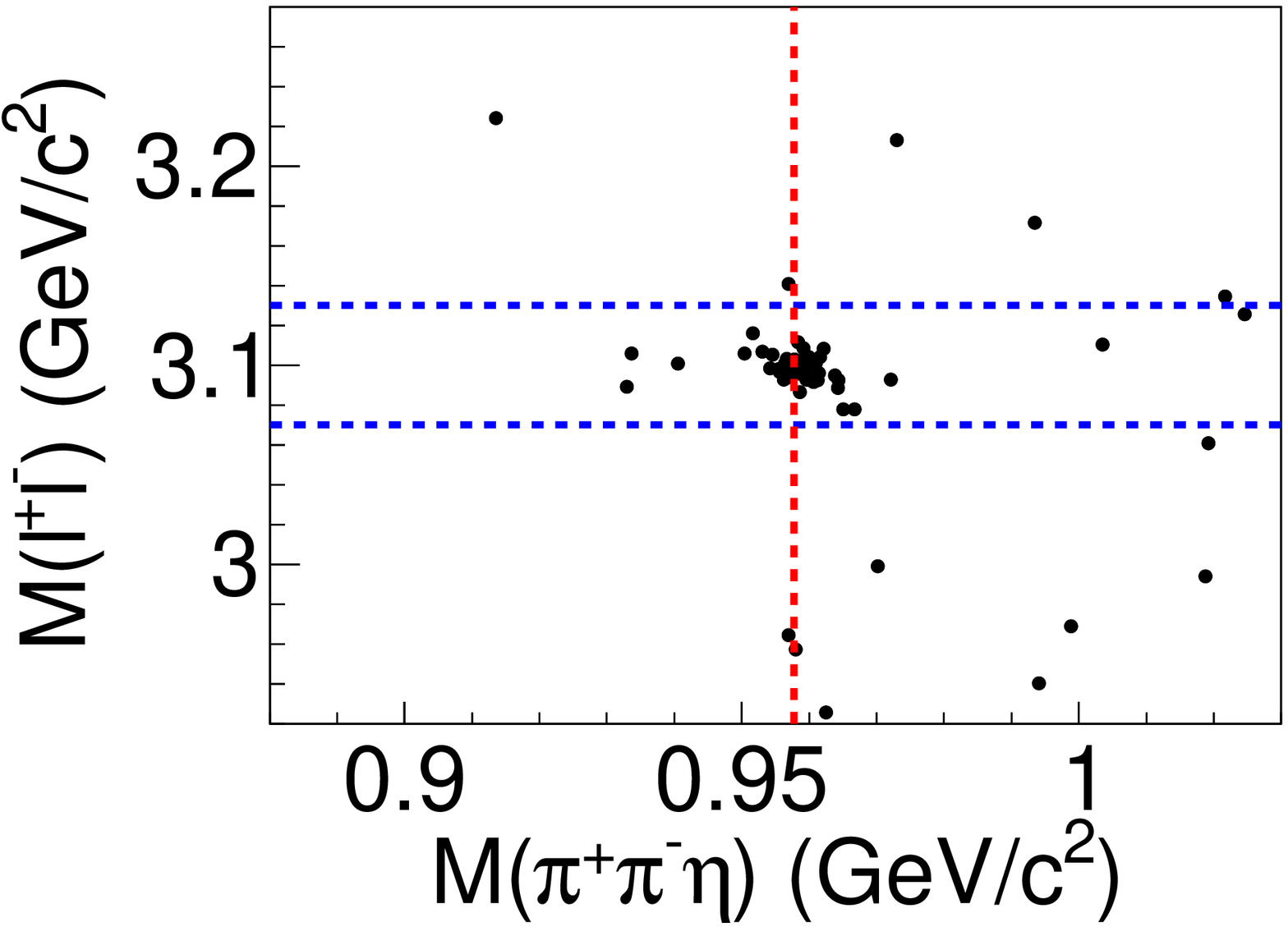}
\put(97,69){(c)}
\end{overpic}
\begin{overpic}[width=0.23\textwidth]{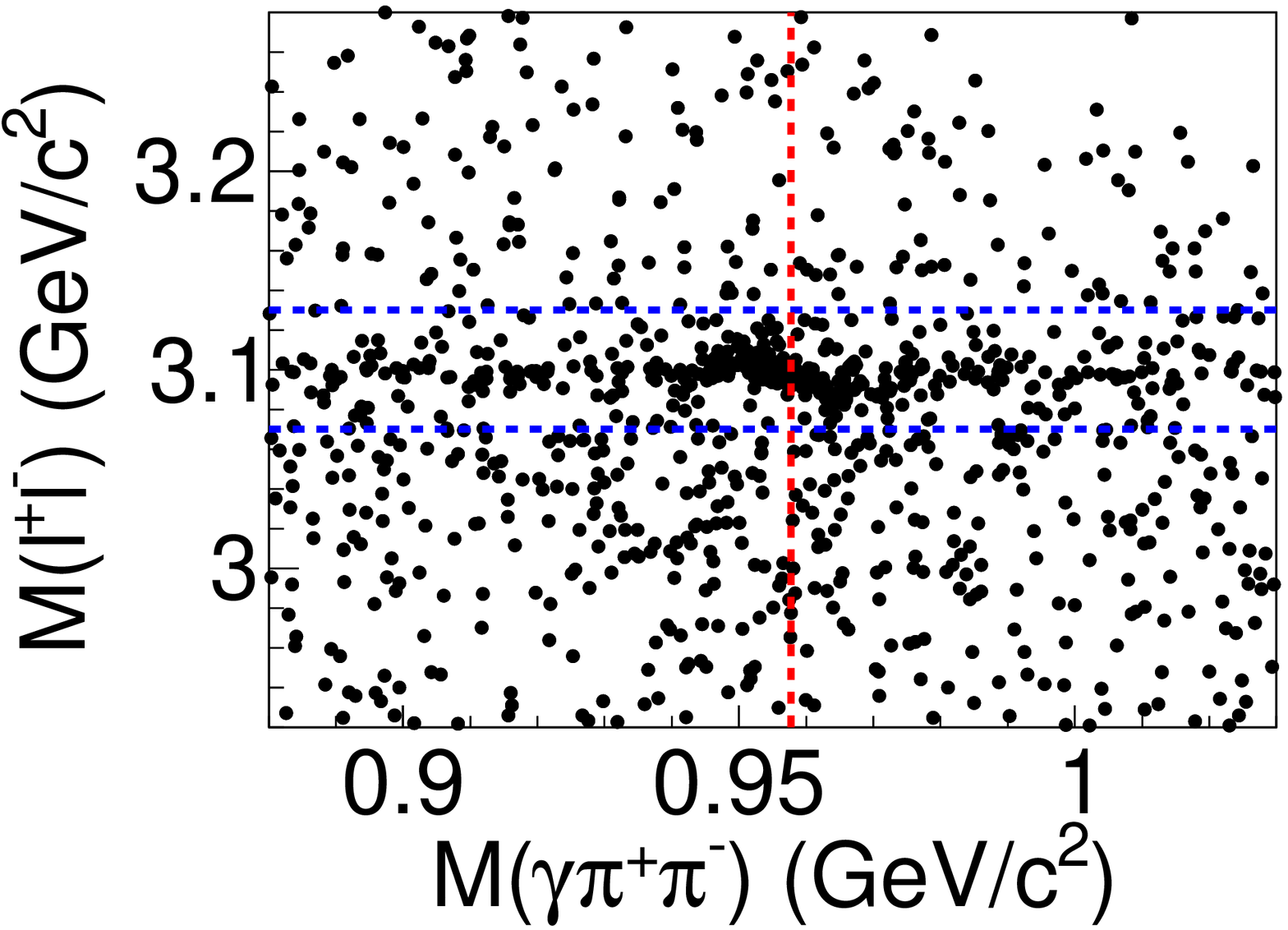}
\put(97,69){(d)}
\end{overpic}
\caption{Distributions of selected events for data at $\sqrt{s}=4.178$ GeV and combined data at other 13 energy points. (a) $M(\ell^+\ell^-)$ versus $M(\pi^{+}\pi^{-}\eta)$ for $\eta'\too\pi^{+}\pi^{-}\eta$ for data at $\sqrt{s}=4.178$ GeV. (b) $M(\ell^+\ell^-)$ versus $M(\gamma\pi^{+}\pi^{-})$ for $\eta'\too\gamma\pi^{+}\pi^{-}$ for data at $\sqrt{s}=4.178$ GeV.
(c) $M(\ell^+\ell^-)$ versus $M(\pi^{+}\pi^{-}\eta)$ for $\eta'\too\pi^{+}\pi^{-}\eta$ for combined data at other 13 energy points. (d) $M(\ell^+\ell^-)$ versus $M(\gamma\pi^{+}\pi^{-})$ for $\eta'\too\gamma\pi^{+}\pi^{-}$ for combined data at other 13 energy points.
The horizontal dashed lines denote the signal region of the $J/\psi$ and the vertical dashed lines mark the nominal $\eta'$ mass.}
\label{fig:scatter}
\end{center}
\end{figure}

Scatter plots of the $\ell^+\ell^-$ invariant mass, $M(\ell^+\ell^-)$, and the $\pi^{+}\pi^{-}\eta/\gamma\pi^{+}\pi^{-}$ invariant masses, $M(\pi^{+}\pi^{-}\eta)/M(\gamma\pi^{+}\pi^{-})$, are shown in Fig.~\ref{fig:scatter} for data taken at $\sqrt{s} = 4.178$ GeV and combined data taken at other 13 energy points. A high-density area can be observed originating from the $\EE \too \eta'J/\psi$ decay. The $J/\psi$ signal region is defined by the mass range [3.07, 3.13]~GeV/$c^{2}$ in $M(\ell^+\ell^-)$ and is indicated by horizontal dashed lines. Sideband regions, defined by the ranges [3.00, 3.06] GeV$/c^{2}$ and [3.14, 3.20] GeV$/c^{2}$, are used to study the non-resonant background. The nominal $\eta'$ mass is indicated by the vertical dashed lines.

\begin{figure}[htbp]
\begin{center}
\begin{overpic}[width=0.23\textwidth]{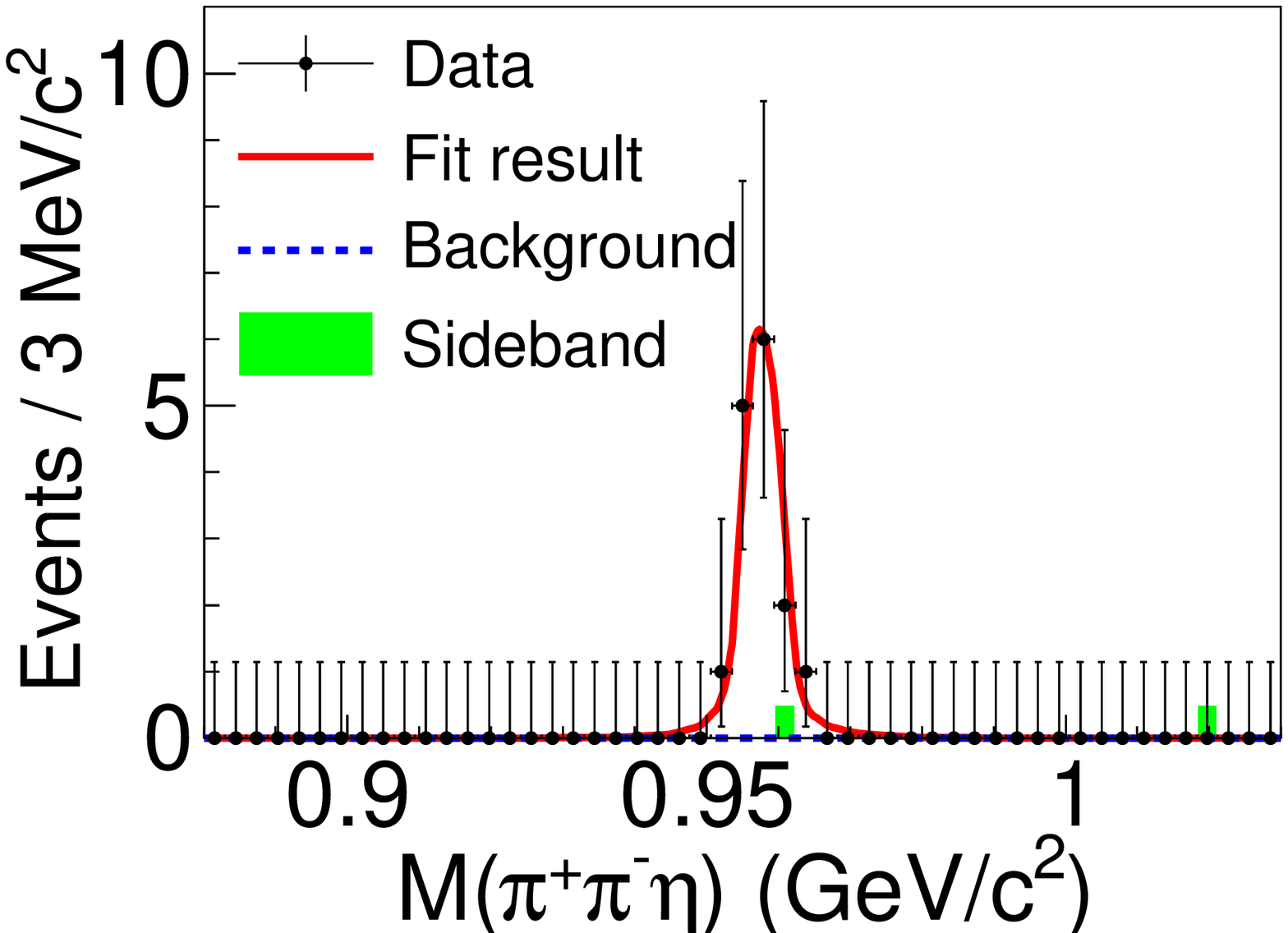}
\put(97,69){(a)}
\end{overpic}
\begin{overpic}[width=0.23\textwidth]{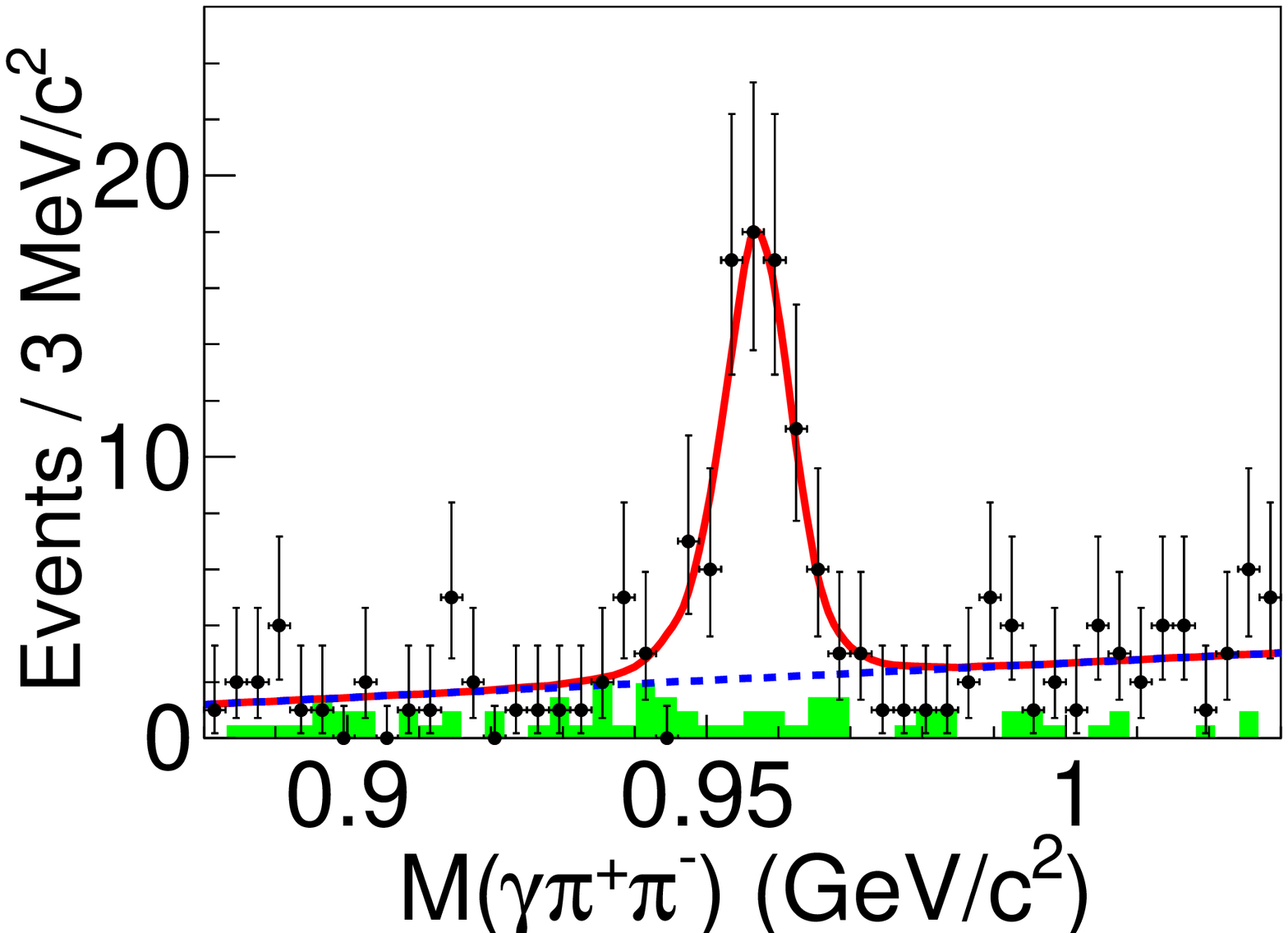}
\put(97,69){(b)}
\end{overpic}
\begin{overpic}[width=0.23\textwidth]{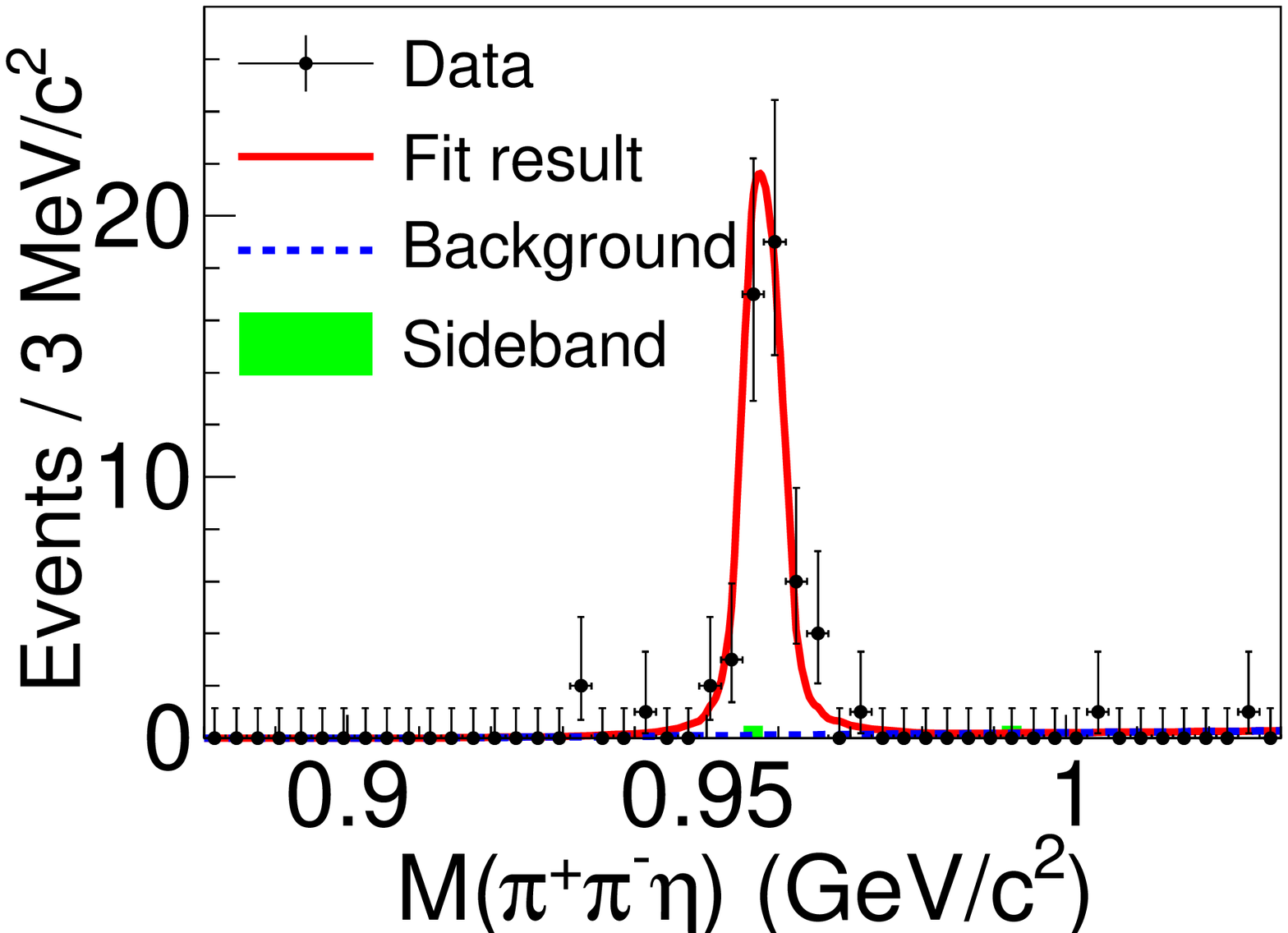}
\put(97,69){(c)}
\end{overpic}
\begin{overpic}[width=0.23\textwidth]{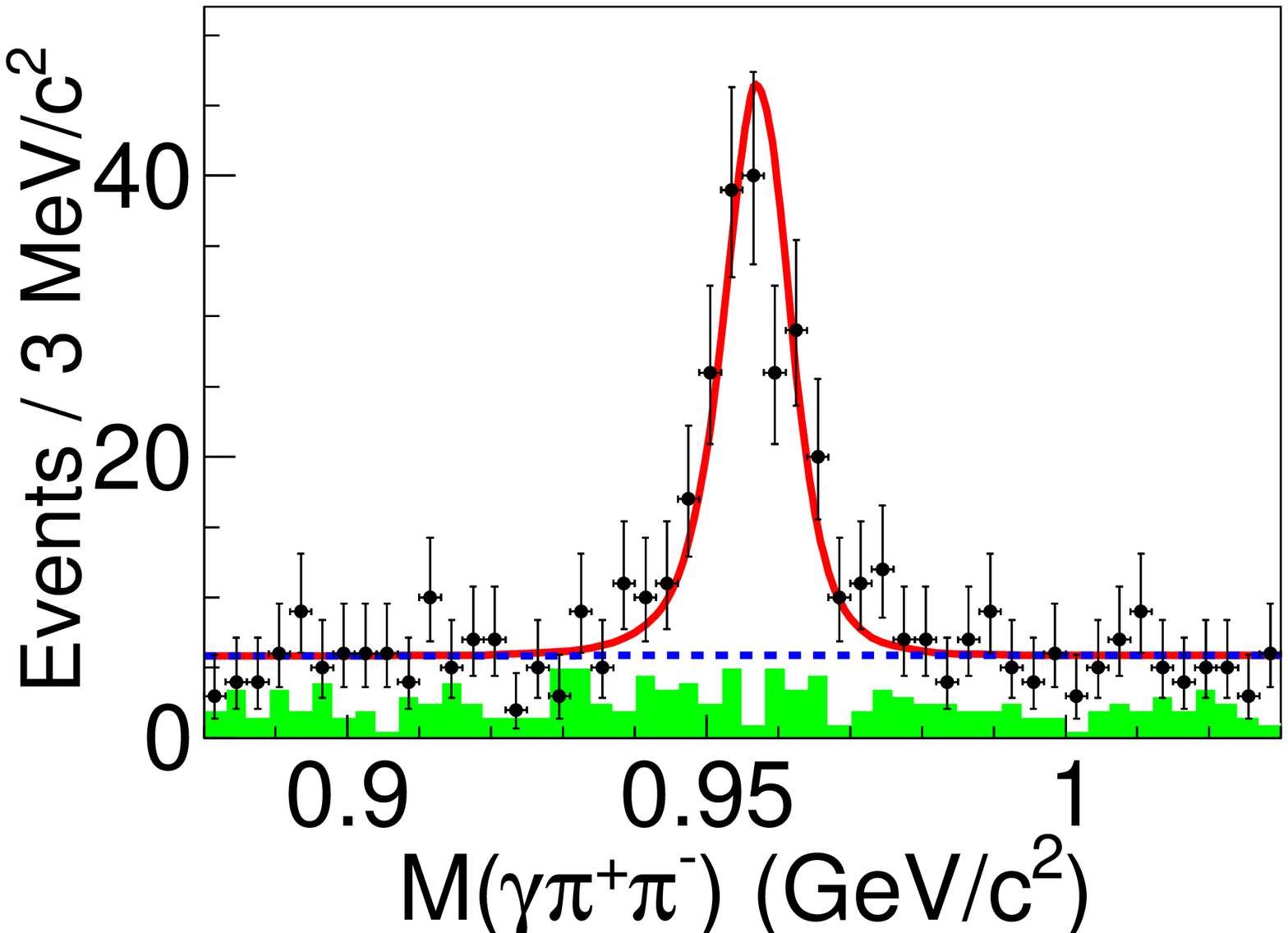}
\put(97,69){(d)}
\end{overpic}
\caption{(Color online) Results of the simultaneous fits to the two invariant mass distributions of $M(\pi^{+}\pi^{-}\eta)$ and $M(\gamma\pi^{+}\pi^{-})$ for data at $\sqrt{s}=4.178$ GeV and combined data at other 13 energy points.
(a) $M(\pi^{+}\pi^{-}\eta)$ for data at $\sqrt{s}=4.178$ GeV. (b) $M(\gamma\pi^{+}\pi^{-})$ for data at $\sqrt{s}=4.178$ GeV.
(c) $M(\pi^{+}\pi^{-}\eta)$ for combined data at other 13 energy points. (d) $M(\gamma\pi^{+}\pi^{-})$ for combined data at other 13 energy points.
The red solid lines are the total fits to data and the blue dashed lines are the background components. The green shaded histograms correspond to the normalized events from the $J/\psi$ sideband region.}
\label{fig:fit}
\end{center}
\end{figure}

Figure~\ref{fig:fit} shows the distributions of $M(\pi^{+}\pi^{-}\eta)$ and $M(\gamma\pi^{+}\pi^{-})$ for data in the $J/\psi$ signal region. Signals for the $\eta'$ meson are observed. The shaded histograms correspond to the normalized events from the $J/\psi$ sideband region. In order to extract the signal yield, a simultaneous maximum likelihood fit is performed to the two $\eta'$ decay modes. The $\eta'$ signal is modeled by the MC-determined shape, and the background is described with a 1st order polynomial.
In the fit, the total signal yield is a free parameter, the ratio of the number of $\eta'\too\pi^{+}\pi^{-}\eta$ signal events to the number of $\eta'\too\gamma\pi^{+}\pi^{-}$ signal events is fixed to
$\frac{\mathcal{B}(\eta' \too \pi^{+}\pi^{-}\eta)\mathcal{B}(\eta \too \gamma\gamma)\epsilon_{\pi^{+}\pi^{-}\eta}}{\mathcal{B}(\eta' \too \gamma\pi^{+}\pi^{-})\epsilon_{\gamma\pi^{+}\pi^{-}}}$,
where $\epsilon_{\pi^{+}\pi^{-}\eta}$ and $\epsilon_{\gamma\pi^{+}\pi^{-}}$ are the efficiencies for the $\pi^{+}\pi^{-}\eta$ and $\gamma\pi^{+}\pi^{-}$ decay modes, respectively.
$\mathcal{B}(\eta' \too \pi^{+}\pi^{-}\eta)$, $\mathcal{B}(\eta \too \gamma\gamma)$ and $\mathcal{B}(\eta' \too \gamma\pi^{+}\pi^{-})$ are the branching fractions, and are taken from PDG~\cite{pdg}.
The solid curves in Fig.~\ref{fig:fit} show the fit results. Data taken at all center-of-mass energies are analyzed using the same method and the fit results are summarized in Table~\ref{tab:crosssection}.

\begin{table*}[htbp]
\begin{center}
  \caption{Born cross sections $\sigma^{\rm B}$ (or upper limits $\sigma^{\rm B}_{\text{upper}}$ at 90\% C.L.) for the reaction $e^+e^-\too\eta'J/\psi$ at different center-of-mass energies $\sqrt{s}$, together with integrated luminosities $\mathcal{L}$, number of signal events $N^{\rm sig}$, radiative correction factors 1 + $\delta$(s), vacuum polarization factors $\frac{1}{|1-\Pi|^{2}}$,
and efficiencies $\epsilon_{\pi^{+}\pi^{-}\eta}$ and $\epsilon_{\gamma\pi^{+}\pi^{-}}$. The first uncertainties are statistical, and the second systematic.}
\label{tab:crosssection}
\begin{tabular}{c  c  c  c  c  c  c  c }
  \hline
  \hline
  \ \ $\sqrt{s}$ (GeV) \ \ & \ \ $\mathcal{L}$ (pb$^{-1}$) \ \ & \ \ $N^{\rm sig}$ \ \ & \ \ 1 + $\delta$(s) \ \ & \ \ $\frac{1}{|1-\Pi|^{2}}$ \ \ & \ \ $\epsilon_{\pi^{+}\pi^{-}\eta}$ $(\%)$ \ \ & \ \ $\epsilon_{\gamma\pi^{+}\pi^{-}}$ $(\%)$ \ \ & \ \ $\sigma^{\rm B}(\sigma^{\rm B}_{\text{upper}})$ (pb) \ \ \\
  \hline
  4.178 & 3194.5 & $86.2\pm10.3$& 0.725 & 1.055 & 15.38 & 33.24 & $2.43\pm0.29\pm0.17$  \\
  4.189 & 524.6  & $13.1\pm4.3$ & 0.739 & 1.056 & 15.57 & 32.94 & $2.21\pm0.73\pm0.17$  \\
  4.199 & 526.0  & $17.6\pm5.0$ & 0.759 & 1.057 & 15.89 & 32.88 & $2.87\pm0.82\pm0.23$  \\
  4.209 & 518.0  & $16.2\pm4.5$ & 0.776 & 1.057 & 15.87 & 31.97 & $2.68\pm0.75\pm0.20$  \\
  4.219 & 514.6  & $14.8\pm4.5$ & 0.783 & 1.057 & 15.95 & 31.65 & $2.46\pm0.75\pm0.19$  \\
  4.226 & 1056.4 & $46.0\pm7.6$ & 0.785 & 1.057 & 16.48 & 32.37 & $3.63\pm0.60\pm0.28$  \\
  4.236 & 530.3  & $18.1\pm5.2$ & 0.799 & 1.056 & 16.38 & 31.72 & $2.85\pm0.82\pm0.21$  \\
  4.244 & 538.1  & $25.0\pm5.8$ & 0.824 & 1.056 & 16.42 & 31.06 & $3.81\pm0.89\pm0.27$  \\
  4.258 & 828.4  & $36.0\pm7.0$ & 0.878 & 1.054 & 16.45 & 30.39 & $3.41\pm0.66\pm0.25$  \\
  4.267 & 531.1  & $19.1\pm4.7$ & 0.914 & 1.053 & 15.64 & 29.16 & $2.83\pm0.70\pm0.21$  \\
  4.278 & 175.7  & $1.0\pm1.0(<3.9)$  & 0.953 & 1.053 & 14.95 & 27.57 & $0.45\pm0.45\pm0.04(<1.77)$  \\
  4.358 & 543.9  & $1.4\pm1.4(<5.0)$  & 1.133 & 1.051 & 12.75 & 22.87 & $0.21\pm0.21\pm0.02(<0.74)$  \\
  4.416 & 1043.9 & $15.3\pm4.5$ & 1.200 & 1.053 & 11.90 & 21.55 & $1.18\pm0.35\pm0.15$  \\
  4.600 & 586.9  & $1.5\pm2.2(<6.2)$  & 1.300 & 1.055 & 10.87 & 19.22 & $0.21\pm0.31\pm0.02(<0.88)$  \\
  \hline
  \hline
\end{tabular}
\end{center}
\end{table*}

The Born cross section is calculated with

\begin{equation}
    \sigma^{\rm B} = \frac{N^{\rm sig}}{\mathcal{L}(1 + \delta (s))\frac{1}{|1-\Pi|^{2}}({\mathcal{B}_{1}}{\epsilon_{\pi^{+}\pi^{-}\eta}}+{\mathcal{B}_{2}}{\epsilon_{\gamma\pi^{+}\pi^{-}}})} ,
\end{equation}
where $N^\text{sig}$ is the total number of signal events, $\mathcal{L}$ is the integrated luminosity obtained using the same method in Ref.~\cite{luminosity}, 1 + $\delta (s)$ is the ISR correction factor obtained from a Quantum Electrodynamics calculation~\cite{QED, KKMC}, $\frac{1}{|1-\Pi|^{2}}$ is the correction factor for vacuum polarization~\cite{vacuum}, $\mathcal{B}_{1}$ is the product of branching fractions $\mathcal{B}(J/\psi\too\ell^+\ell^-)\times\mathcal{B}(\eta'\too\pi^{+}\pi^{-}\eta)\times\mathcal{B}(\eta\too\gamma\gamma)$,
and $\mathcal{B}_{2}$ is the product of branching fractions $\mathcal{B}(J/\psi\too\ell^+\ell^-)\times\mathcal{B}(\eta'\too\gamma\pi^{+}\pi^{-})$.
For data at $\sqrt{s}=4.278$, $4.358$ and $4.600$ GeV, which have no significant signals, we calculate upper limits at $90\%$ confidence level (C.L.) using the Bayesian method assuming a uniform prior distribution. The upper limit on the number of $\eta'$ signal events $N^{\text{up}}_{\eta'}$ at $90\%$ C.L. is obtained by solving the equation $\int_{0}^{N^{\text{up}}_{\eta'}}F(x)dx/\int_{0}^{\infty}F(x)dx=0.90$, where $F(x)$ is the posterior distribution (of signal events), which is the likelihood function multiplied by the prior distribution. The systematic uncertainty is taken into account by smearing the posterior distribution. The Born cross sections (or upper limits at 90\% C.L.) at each energy point for
$\EE\too\eta'J/\psi$ are listed in Table~\ref{tab:crosssection}. The efficiencies $\epsilon_{\pi^{+}\pi^{-}\eta}$ and $\epsilon_{\gamma\pi^{+}\pi^{-}}$ in Table~\ref{tab:crosssection} are rapidly decreasing above 4.26 GeV, they are due to the ISR correction effect.

\section{V. systematic uncertainty}

The systematic uncertainties of the Born cross section measurement originate from the luminosity determination, the tracking efficiency, the photon detection efficiency, the kinematic fit, the $J/\psi$ mass window, the radiative correction, the fit range, the signal and the background modeling, and the input branching fractions.

The luminosities are measured with a precision of $1.0\%$ using the Bhabha process~\cite{luminosity}.
The uncertainty in the tracking efficiency is $1.0\%$ per track~\cite{omegachic2}. Since the two decay channels have the same number of charged tracks in the same region of momenta, their tracking efficiencies are fully correlated.  Therefore, a 4.0\% uncertainty is introduced to the final results.

The uncertainty in photon reconstruction is $1.0\%$ per photon~\cite{photon}. There are two photons for the $\eta' \too \pi^{+}\pi^{-}\eta$ mode and one photon for $\eta' \too \gamma\pi^{+}\pi^{-}$.
Therefore, we vary the values $\epsilon_{\pi^{+}\pi^{-}\eta}$ and $\epsilon_{\gamma\pi^{+}\pi^{-}}$ up or down by $1\%\times N_{\gamma}$ and refit the data, where $N_{\gamma}$ is the number of photons in the final state.
The maximum change of the measured cross section is taken as the systematic uncertainty.

The uncertainty due to the kinematic fit is estimated by correcting the helix parameters of charged tracks according to the method described in Ref.~\cite{helix}. The difference between detection efficiencies obtained from MC samples with and without this correction is taken as the uncertainty.

The uncertainty for the $J/\psi$ mass window requirement is estimated using $\EE \too \gamma_{ISR}\psi(3686), \psi(3686) \too \pi^{+}\pi^{-}J/\psi$ events. The difference of efficiency between data and MC simulation is found to be 1.6$\%$~\cite{jpsimasswindow}.

The line shape of the $\EE \too \eta' J/\psi$ cross section will affect the radiative correction factor and the efficiency. In the nominal results, we use a coherent sum of $\psi(4160)$ and $\psi(4260)$ resonances~\cite{pdg} as the line shape. To estimate the uncertainty from the radiative correction, we change the line shape to a coherent sum of $\psi(4160)$, $\psi(4260)$, and $\psi(4415)$ resonances, a coherent sum of $\psi(4160)$, $Y(4220)$, and $Y(4320)$ resonances~\cite{pipijpsi-bes}, a coherent sum of $\psi(4160)$, $\psi(4260)$, and a continuum component, and take the largest difference of the cross section measurement to the nominal one as the systematic uncertainty.

Due to limited statistics, we add all data together to estimate the uncertainties from the fit range, the signal shape, and the background shape. The uncertainty from the fit range is obtained by varying the boundary of the fit range by $\pm$0.01 GeV/$c^{2}$. We take the largest difference of the cross section measurement to the nominal one as the systematic uncertainty. For the uncertainty from the signal shape, we use the MC-determined shape convolved with a Gaussian function to refit the data. The Gaussian function compensates for a possible mass resolution discrepancy between data and MC simulations, and its parameters are free. The systematic uncertainty due to the background shape is estimated by changing the background shape from a 1st-order polynomial to a 2nd-order polynomial, and taking the difference as the uncertainty.
The uncertainties from the input branching fractions are taken from PDG~\cite{pdg}.

Table~\ref{tab:sumerror} summarizes all the systematic uncertainties related to the cross section measurement of the $\EE \too \eta' J/\psi$ process for each center-of-mass energy. The overall systematic uncertainties are obtained by adding all the sources of systematic uncertainties in quadrature assuming they are uncorrelated.
\begin{table*}[htbp]
\caption{Relative systematic uncertainties (in $\%$) from the different sources.}
\label{tab:sumerror}
\begin{tabular}{c c c c c c c c c c c c c c c}
  \hline
  \hline
\ Source / $\sqrt{s}$ (GeV) \ & \ 4.178 \ & \ 4.189 \ & \ 4.199 \ & \ 4.209 \ & \ 4.219 \ & \ 4.226 \ & \ 4.236 \ & \ 4.244 \ & \ 4.258 \ & \ 4.267 \ & \ 4.278 \ & \ 4.358 \ & \ 4.416 \ & \ 4.600 \ \\
  \hline
  Luminosity & 1.0 & 1.0 & 1.0 & 1.0 & 1.0 & 1.0 & 1.0 & 1.0 & 1.0 & 1.0 & 1.0 & 1.0 & 1.0 & 1.0 \\
  Tracking efficiency & 4.0 & 4.0 & 4.0 & 4.0 & 4.0 & 4.0 & 4.0 & 4.0 & 4.0 & 4.0 & 4.0 & 4.0 & 4.0 & 4.0 \\
  Photon detection & 1.3 & 1.3 & 1.3 & 1.3 & 1.3 & 1.3 & 1.3 & 1.3 & 1.3 & 1.3 & 1.3 & 1.3 & 1.3 & 1.3 \\
  Kinematic fit & 2.7 & 3.0 & 2.8 & 3.2 & 2.8 & 2.5 & 2.8 & 2.9 & 2.9 & 3.2 & 2.9 & 2.9 & 2.7 & 2.8  \\
  $J/\psi$ mass window & 1.6 & 1.6 & 1.6 & 1.6 & 1.6 & 1.6 & 1.6 & 1.6 & 1.6 & 1.6 & 1.6 & 1.6 & 1.6 & 1.6 \\
  Radiative correction & 1.2 & 3.0 & 3.5 & 2.2 & 3.1 & 3.6 & 1.5 & 0.9 & 1.3 & 2.1 & 4.6 & 7.8 & 10.6 & 1.3 \\
  Fit Range & 0.3 & 0.3 & 0.3 & 0.3 & 0.3 & 0.3 & 0.3 & 0.3 & 0.3 & 0.3 & 0.3 & 0.3 & 0.3 & 0.3 \\
  Signal shape & 2.6 & 2.6 & 2.6 & 2.6 & 2.6 & 2.6 & 2.6 & 2.6 & 2.6 & 2.6 & 2.6 & 2.6 & 2.6 & 2.6 \\
  Background shape & 3.1 & 3.1 & 3.1 & 3.1 & 3.1 & 3.1 & 3.1 & 3.1 & 3.1 & 3.1 & 3.1 & 3.1 & 3.1 & 3.1 \\
  Branching fraction & 1.6 & 1.6 & 1.6 & 1.6 & 1.6 & 1.6 & 1.6 & 1.6 & 1.6 & 1.6 & 1.6 & 1.6 & 1.6 & 1.6 \\
  Sum & 7.0 & 7.6 & 7.8 & 7.4 & 7.6 & 7.7 & 7.1 & 7.0 & 7.1 & 7.4 & 8.4 & 10.5 & 12.6 & 7.1   \\
  \hline
  \hline
\end{tabular}
\end{table*}

\section{VI. discussion}
\begin{figure}[htbp]
\begin{center}
\begin{overpic}[width=0.43\textwidth]{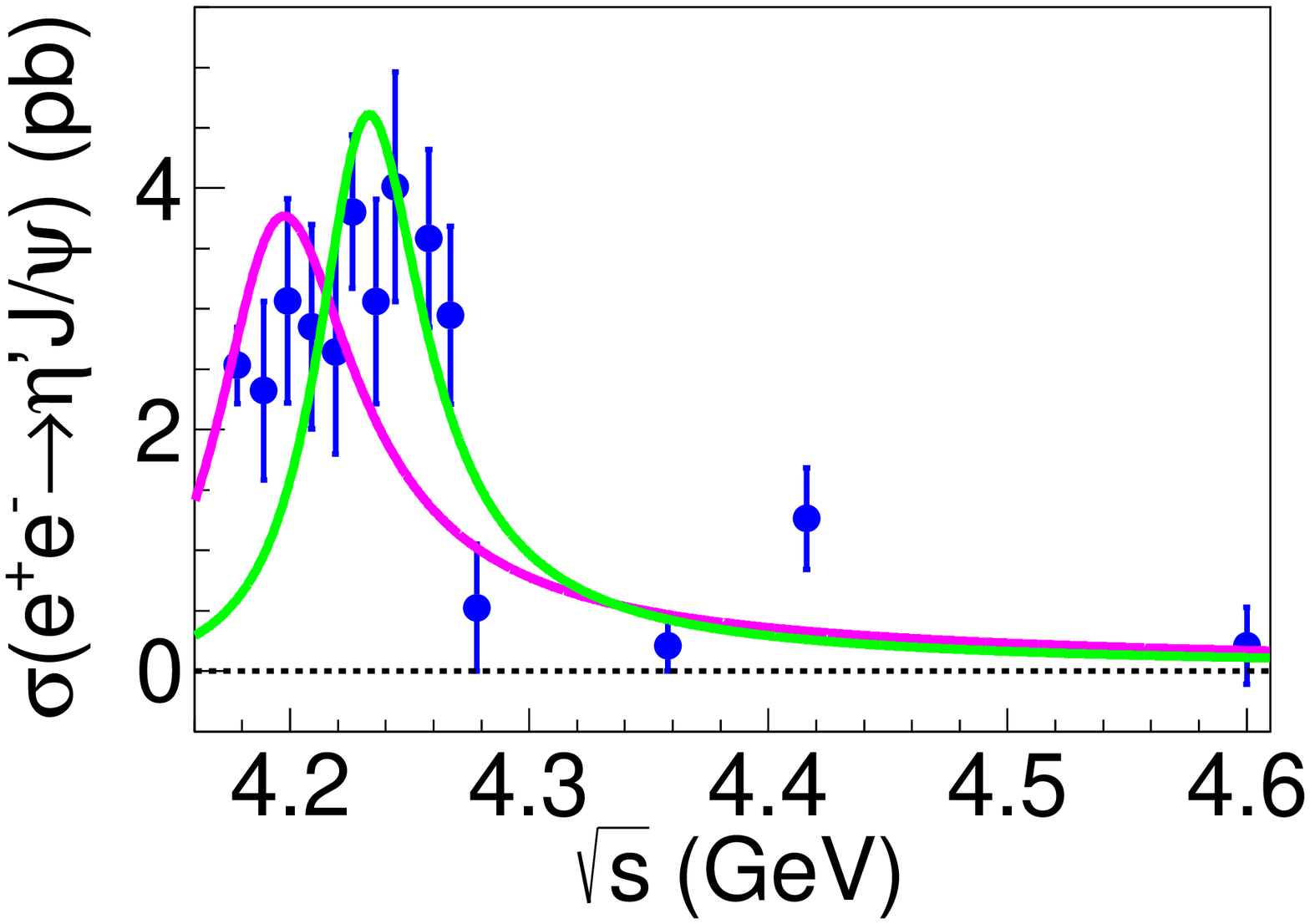}
\put(182,129){\LARGE{(a)}}
\end{overpic}
\begin{overpic}[width=0.43\textwidth]{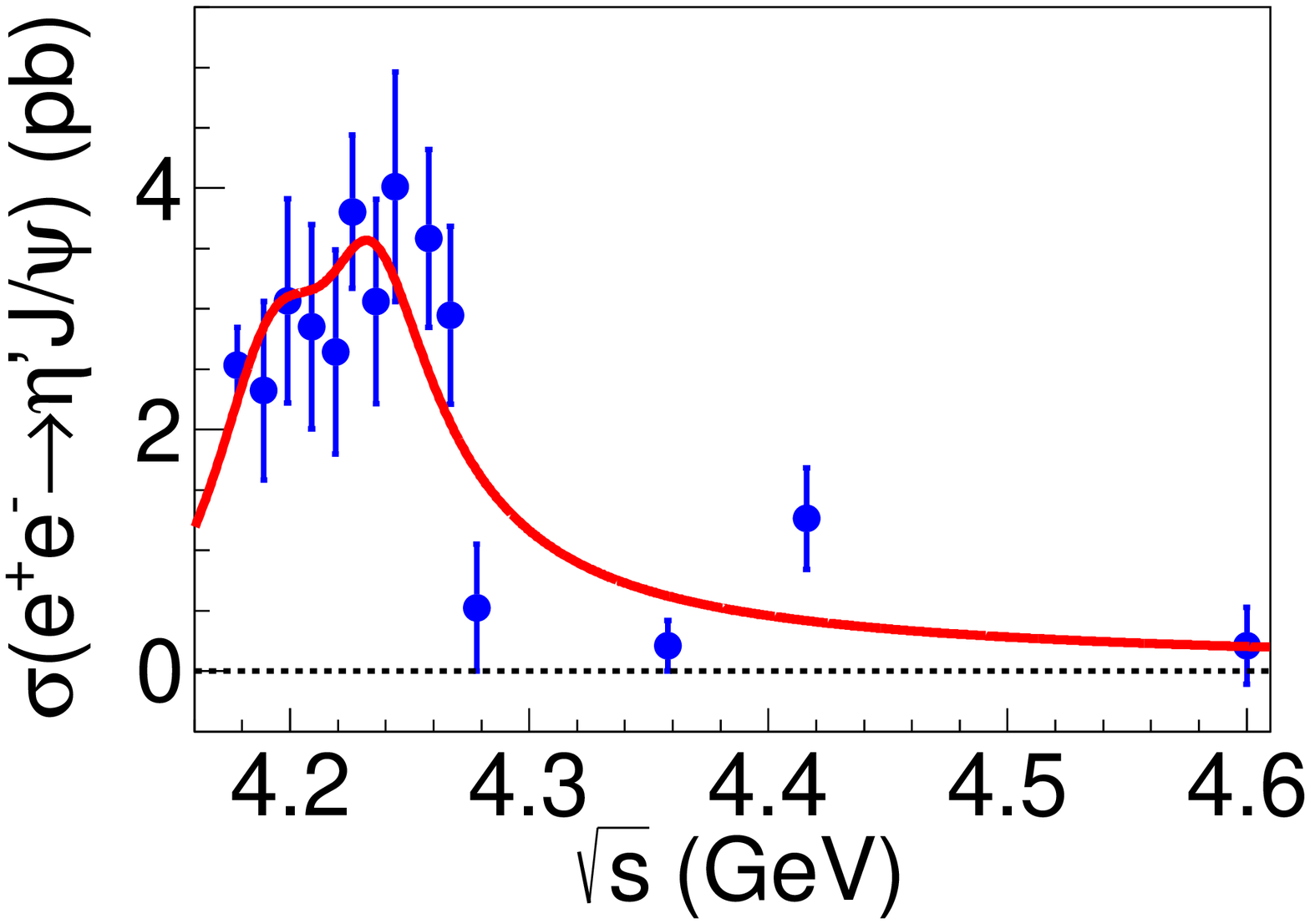}
\put(182,129){\LARGE{(b)}}
\end{overpic}
\caption{(a) Fit to the $\EE \too \eta'J/\psi$ cross section with a single $\psi(4160)$ resonance (pink solid line) or a single $\psi(4260)$ resonance (green solid line). (b) Fit to the $\EE \too \eta'J/\psi$ cross section with a coherent sum of $\psi(4160)$ and $\psi(4260)$ resonances (red solid line).}
\label{fig:crosssection}
\end{center}
\end{figure}

Figure~\ref{fig:crosssection} shows the dressed cross sections ($\sigma=\frac{\sigma^{\rm B}}{|1-\Pi|^2}$) for the $\EE \too \eta'J/\psi$ reaction at different energy points. We observe an enhancement in the cross section around 4.2 GeV. By assuming that the $\eta'J/\psi$ signals come from a single resonance $\psi(4160)$ or $\psi(4260)$, with mass $M$ and width $\Gamma$ that are fixed to their PDG values~\cite{pdg}, we use a least $\chi^2$ method to fit the cross section data with the following formula :

\begin{equation}
    \sigma(\sqrt{s}) = |\frac{\sqrt{12\pi\Gamma_{ee}\mathcal{B}(\eta' J/\psi)\Gamma}}{s-M^{2}+iM\Gamma}\sqrt{\frac{\Phi^{3}(\sqrt{s})}{\Phi^{3}(M)}}|^{2},
\end{equation}
where $\Phi(\sqrt{s})=p/\sqrt{s}$ is the two-body phase space factor, $p$ is the $\eta'$ momentum in the $\EE$ center-of-mass frame, and $\Gamma_{ee}$ is the electronic width of the $\psi(4160)$ or $\psi(4260)$. The $\chi^2$ function is constructed as:

\begin{equation}
    \chi^2 = \sum_{i=1}^{n}\frac{(\sigma_{i}^{\text{data}}-\sigma_{i}^{\text{fit}})^2}{\Delta_{i}^2},
\end{equation}
where $\sigma_{i}^{\text{data}}$ and $\sigma_{i}^{\text{fit}}$ are the measured and fitted cross sections of the $i$-th energy point, respectively, and $\Delta_{i}$ is the corresponding statistical uncertainty. The goodness of fit is $\chi^2/\text{NDF}=38/13$, corresponding to a confidence level of $2.9\times10^{-4}$, for a single resonance $\psi(4160)$ and $\chi^2/\text{NDF}=63/13$, corresponding to a confidence level of $1.5\times10^{-8}$, for a single resonance $\psi(4260)$,
where $\text{NDF}$ is the number of degrees of freedom. The fit qualities indicate that the data cannot be described well by a single $\psi(4160)$ or $\psi(4260)$ resonance.

Then we try to use a coherent sum of $\psi(4160)$ and $\psi(4260)$ resonances to fit the $\EE \too \eta'J/\psi$ cross section, where the resonances' parameters are fixed to those from PDG~\cite{pdg}. The fit result is shown in Fig.~\ref{fig:crosssection} and Table~\ref{tab:fitresulta}. The goodness of fit is $\chi^2/\text{NDF}=19/11$, corresponding to a confidence level of $6.1\%$, indicating that the $\EE \too \eta'J/\psi$
cross section can be described by a coherent sum of $\psi(4160)$ and $\psi(4260)$.
The significances for the $\psi(4160)$ and $\psi(4260)$ are $6.3\sigma$ and $4.0\sigma$. Significance of $\psi(4160)$ is to compare to single $\psi(4260)$ fit, while vice versa. In additional, we try to use a coherent sum of $\psi(4160)$, $Y(4220)$ and $Y(4320)$ resonances to fit, where $Y(4220)$ and $Y(4320)$'s parameters are fixed to the results in Ref.~\cite{pipijpsi-bes}. The goodness of fit is $\chi^2/\text{NDF}=14/9$, corresponding to a confidence level of $12.2\%$. The contribution of the continuum process is studied by means of a phase space function $\Phi^{3}(\sqrt{s})$ or a $\frac{1}{s}$ parametrization, and the cross section is fitted again taking into account this additional factor. We find that the additional contribution of the continuum is not statistically significant. We also try to use one BW function to fit the cross section, the fitted mass and width are $M=(4200\pm6)$ MeV/$c^2$ and $\Gamma=(89\pm11)$ MeV, the goodness of the fit is $\chi^2/\text{NDF}=26/11$, corresponding to a confidence level of $6.5\times10^{-3}$.

\begin{table}[htbp]
\begin{center}
\caption{ The fitted parameters of the cross section of $\EE \too \eta'J/\psi$ using a coherent sum of $\psi(4160)$ and $\psi(4260)$. ``Solution I" represents the constructive solution, and ``Solution II" represents the destructive solution. The uncertainty is statistical only.}
\label{tab:fitresulta}
\begin{tabular}{ccc}
  \hline
  \hline
  Parameter & Solution I & Solution II \\
  \hline
  $\Gamma^{\psi(4160)}_{ee}\mathcal{B}(\psi(4160)\too\eta'J/\psi)$ (eV) & $0.17\pm0.04$ & $1.07\pm0.09$ \\
  $\Gamma^{\psi(4260)}_{ee}\mathcal{B}(\psi(4260)\too\eta'J/\psi)$ (eV) & $0.06\pm0.03$ & $1.38\pm0.11$ \\
  $\phi$ (rad) & $-0.03\pm0.44$ & $2.54\pm0.04$ \\
  \hline
  \hline
\end{tabular}
\end{center}
\end{table}

\section{VII. Summary}
The process $\EE \too \eta'J/\psi$ has been studied using 14 data samples collected at center-of-mass energies from $\sqrt{s} =$ 4.178 to 4.600 GeV. The $\sqrt{s}$-dependence of the cross section has been measured. In the previous study, the process $\EE \too \eta'J/\psi$ was only observed at $\sqrt{s}$ = 4.226 and 4.258~GeV, which is not sufficient to constrain the parameterization of the line shape of $\EE \too \eta'J/\psi$ around $\sqrt{s} =$ 4.2 GeV. In this study, the cross section of $\EE \too \eta'J/\psi$ is measured by adding more data samples at nine energy points in the range 4.178 $\leqslant\sqrt{s}\leqslant$ 4.278 GeV,
which improves our understanding of the line shape of $\EE \too \eta'J/\psi$ around $\sqrt{s} =$ 4.2 GeV. The results of the data samples at the previous five energy points are also updated. The $\EE \too \eta'J/\psi$ cross section cannot be properly described by a single $\psi(4160)$ or $\psi(4260)$ resonance, while a coherent sum of $\psi(4160)$ and $\psi(4260)$ offers a better description. Further experimental studies with higher statistics are needed to draw a clearer conclusion on the structures in the $\EE \too \eta'J/\psi$ process.
The cross section of $\EE \too \eta'J/\psi$ is about an order of magnitude lower than that of $\EE \too \eta J/\psi$~\cite{bes}, and the line shape of $\EE \too \eta'J/\psi$ is relatively flat from $\sqrt{s}$ = 4.2 to 4.26~GeV, while that of $\EE \too \eta J/\psi$ is sharply drop. The precise measurements of $\EE \too \eta'J/\psi$ and $\eta J/\psi$ in the future may be useful inputs for a study of $\eta-\eta'$ mixing.

\section{ACKNOWLEDGMENTS}
The BESIII collaboration thanks the staff of BEPCII and the IHEP computing center for their strong support. This work is supported in part by National Key Basic Research Program of China under Contract No. 2015CB856700; National Natural Science Foundation of China (NSFC) under Contracts Nos. 11905179, 11625523, 11635010, 11735014, 11822506, 11835012; the Chinese Academy of Sciences (CAS) Large-Scale Scientific Facility Program; Joint Large-Scale Scientific Facility Funds of the NSFC and CAS under Contracts Nos. U1532257, U1532258, U1732263, U1832207; CAS Key Research Program of Frontier Sciences under Contracts Nos. QYZDJ-SSW-SLH003, QYZDJ-SSW-SLH040; 100 Talents Program of CAS; INPAC and Shanghai Key Laboratory for Particle Physics and Cosmology; ERC under Contract No. 758462; Foundation of Henan Educational Committee under Contract No. 19A140015; Nanhu Scholars Program for Young Scholars of Xinyang Normal University; German Research Foundation DFG under Contracts Nos. Collaborative Research Center CRC 1044, FOR 2359; Istituto Nazionale di Fisica Nucleare, Italy; Koninklijke Nederlandse Akademie van Wetenschappen (KNAW) under Contract No. 530-4CDP03; Ministry of Development of Turkey under Contract No. DPT2006K-120470; National Science and Technology fund; STFC (United Kingdom); The Knut and Alice Wallenberg Foundation (Sweden) under Contract No. 2016.0157; The Royal Society, UK under Contracts Nos. DH140054, DH160214; The Swedish Research Council; U. S. Department of Energy under Contracts Nos. DE-FG02-05ER41374, DE-SC-0010118, DE-SC-0012069; University of Groningen (RuG) and the Helmholtzzentrum fuer Schwerionenforschung GmbH (GSI), Darmstadt.


\begin{thebibliography}{**}

\bibitem{etapupsilon} E.~Guido {\it et al.} [Belle Collaboration], Phys.\ Rev.\ Lett.\ {\bf 121}, 062001 (2018).

\bibitem{cleo} T.~E.~Coan {\it et al.} [CLEO Collaboration], Phys.\ Rev.\ Lett.\ {\bf 96}, 162003 (2006).

\bibitem{bes} M.~Ablikim {\it et al.} [BESIII Collaboration], Phys.\ Rev.\ D {\bf 86}, 071101 (2012); Phys.\ Rev.\ D {\bf 91}, 112005 (2015).

\bibitem{belle} X.~L.~Wang {\it et al.} [Belle Collaboration], Phys.\ Rev.\ D {\bf 87}, 051101 (2013).

\bibitem{theory} C.~F.~Qiao and R.~L.~Zhu, Phys.\ Rev.\ D {\bf 89}, 074006 (2014).

\bibitem{etapjpsi} M.~Ablikim {\it et al.} [BESIII Collaboration], Phys.\ Rev.\ D {\bf 94}, 032009 (2016).

\bibitem{omegachic0-bes} M.~Ablikim {\it et al.} [BESIII Collaboration], Phys.\ Rev.\ D {\bf 99}, 091103 (2019).

\bibitem{pipijpsi-bes} M.~Ablikim {\it et al.} [BESIII Collaboration], Phys.\ Rev.\ Lett.\ {\bf 118}, 092001 (2017).

\bibitem{pipihc-bes} M.~Ablikim {\it et al.} [BESIII Collaboration], Phys.\ Rev.\ Lett.\ {\bf 118}, 092002 (2017).

\bibitem{pipipsip-bes} M.~Ablikim {\it et al.} [BESIII Collaboration], Phys.\ Rev.\ D {\bf 96}, 032004 (2017).

\bibitem{piDDstar-bes} M.~Ablikim {\it et al.} [BESIII Collaboration], Phys.\ Rev.\ Lett.\ {\bf 122}, 102002 (2019).

\bibitem{besiii} M.~Ablikim {\it et al.} [BESIII Collaboration], Nucl.\ Instrum.\ Meth.\ A {\bf 614}, 345 (2010).

\bibitem{gammapipi} M.~Ablikim {\it et al.} [BESIII Collaboration], Phys.\ Rev.\ Lett.\ {\bf 120}, 242003 (2018).

\bibitem{bepcii} C.~H.~Yu {\it et al.}, Proceedings of IPAC2016, Busan, Korea, 2016, doi:10.18429/JACoW-IPAC2016-TUYA01.

\bibitem{etof} X.~Li {\it et al.}, Rad. Det. Tech. Meth. {\bf 1}, 13 (2017); Y.~X.~Guo {\it et al.}, Rad. Det. Tech. Meth. {\bf 1}, 15 (2017).

\bibitem{geant4} S.~Agostinelli {\it et al.} [GEANT4 Collaboration], Nucl.\ Instrum.\ Meth.\ A {\bf 506}, 250 (2003).

\bibitem{KKMC} S.~Jadach, B.~F.~L.~Ward and Z.~Was, Phys.\ Rev.\ D {\bf 63}, 113009 (2001); Comput.\ Phys.\ Commun.\  {\bf 130}, 260 (2000).

\bibitem{pdg} M.~Tanabashi {\it et al.} [Particle Data Group], Phys.\ Rev.\ D {\bf 98}, 030001 (2018).

\bibitem{ref:evtgen} D.~J.~Lange, Nucl.\ Instrum.\ Meth.\ A {\bf 462}, 152 (2001); R.~G.~Ping, Chin. Phys. C {\bf 32}, 599 (2008).

\bibitem{ref:lundcharm} J.~C.~Chen, G.~S.~Huang, X.~R.~Qi, D.~H.~Zhang and Y.~S.~Zhu, Phys.\ Rev.\ D {\bf 62}, 034003 (2000); R.~L.~Yang, R.~G.~Ping and H.~Chen, Chin.\ Phys.\ Lett.\  {\bf 31}, 061301 (2014).

\bibitem{photos} E.~Richter-Was, Phys.\ Lett.\ B {\bf 303}, 163 (1993).

\bibitem{luminosity} M.~Ablikim {\it et al.} [BESIII Collaboration], Chin.\ Phys.\ C {\bf 39}, 093001 (2015).

\bibitem{QED} E.~A.~Kuraev and V.~S.~Fadin, Sov.\ J.\ Nucl.\ Phys.\  {\bf 41}, 466 (1985).

\bibitem{vacuum} S.~Actis {\it et al.} [Working Group on Radiative Corrections and Monte Carlo Generators for Low Energies], Eur.\ Phys.\ J.\ C {\bf 66}, 585 (2010).

\bibitem{omegachic2} M.~Ablikim {\it et al.} [BESIII Collaboration], Phys.\ Rev.\ D {\bf 93}, 011102 (2016).

\bibitem{photon} M.~Ablikim {\it et al.} [BESIII Collaboration], Phys.\ Rev.\ D {\bf 99}, 051101 (2019); Phys.\ Rev.\ Lett.\ {\bf 118}, 221802 (2017).

\bibitem{helix} M.~Ablikim {\it et al.} [BESIII Collaboration], Phys.\ Rev.\ D {\bf 87}, 012002 (2013).

\bibitem{jpsimasswindow} M.~Ablikim {\it et al.} [BESIII Collaboration], Phys.\ Rev.\ Lett.\ {\bf 112}, 092001 (2014).

\end{thebibliography}
\end{document}